\documentstyle[aps,prl,epsfig]{revtex}

\begin{document}

\twocolumn[\hsize\textwidth\columnwidth
\hsize\csname@twocolumnfalse\endcsname
\draft

\title{Phonon-assisted tunneling in the quantum regime of
 Mn$_{12}$-ac}

\author{I.  Chiorescu$^1$, R. Giraud$^1$,  A. G. M. Jansen$^2$, A. 
Caneschi$^3$, and B. Barbara$^1$}

\address{$^1$Laboratoire de Magn\'etisme Louis N\'eel CNRS, BP166, 38042
Grenoble, Cedex-09, France\\
$^2$Laboratoire des Champs Magn\'etiques Intenses, GHMFL MPI-CNRS, BP166, 38042
Grenoble, Cedex-09, France\\
$^3$Department of Chemistry, University of Florence, 50144, Italy }

\date{\today}

\maketitle

\begin{abstract}
Longitudinal or transverse magnetic fields applied on a crystal of
Mn$_{12}$-ac allows to observe independent  tunnel transitions between m=-S+p
and m=S-n-p (n=6-10, p=0-2 in longitudinal field and n=p=0 in transverse
field). We observe a smooth transition (in longitudinal) from coherent
ground-state to thermally activated tunneling. Furthermore two ground-state
relaxation regimes showing a crossover between quantum spin relaxation far from
equilibrium and near equilibrium, when the environment destroys multimolecule
correlations. Finally, we stress that the complete Hamiltonian of  Mn$_{12}$
should contain odd spin operators of low order.

\end{abstract}

]

\narrowtext

The system Mn$_{12}$-ac is constituted of magnetic molecules in each of which
12 Mn ions (8 spins-2 and 4 spins-3/2) are strongly coupled by
super-exchange interactions~\cite{Lis}. The resulting spin $S=10$, containing
10$^3$ atoms of different  species (Mn, O, C, H) has volume $\approx1$nm$^3$.
The Hilbert space dimension $D_H$ describing the entangled spin states of the
molecule is huge: $D_H=10^8$.  In the mesoscopic approach to Mn$_{12}$-ac
most observations can be explained with $D_H=2S+1=21$ (for a review,
see~\cite{Igor}). The large energy barrier $DS_z^2$ gives extremely small
ground-state splitting. The consequence is that quantum relaxation is slow and
strongly influenced by fluctuations of the environment. Above 1.5 K a single
crystal of Mn$_{12}$-ac shows staircase-like hysteresis loops when the
magnetic field is applied along the $c-$easy axis of magnetization~\cite{Luc}
(for powder experiments see \cite{BBjmmm95,Paulsen2,Leon,Fried}). Tunneling
rates, deduced from relaxation and ac-susceptibility experiments give sharp
maxima precisely at the fields where hysteresis loops are steep:
$B_n\approx0.44n$ (in Teslas), with $n = 0, 1, 2,\dots$ All these results
constituted the first evidence for thermally assisted tunneling in large spin
molecules. Later we studied the effects of spin and phonon baths in this
tunneling regime and the passage to the classically activated regime
(\cite{LucLowT,jmmm200} and \cite{V15} for a low spin molecule).

In this letter we give a similar study, but at lower temperatures. We determine
the nature of the cross-over between thermally assisted and ground-state
tunneling and show an interesting quantum behaviour of the molecule spin
dynamics, related to spin-phonon transitions. Furtheremore, we suggest that odd
spin operators~\cite{Kahan}, consistent with the tetragonal $S_4$ point
symmetry of Mn$_{12}$-ac~\cite{Lis} should be taken into account in the
Hamiltonian of this system. The magnetization of a small single crystal of
Mn$_{12}$-ac in high fields and at sub-Kelvin temperatures was obtained from
magnetic torque experiments, performed at the Grenoble High Magnetic Field
Laboratory. The way to extract the longitudinal or transverse magnetization
(parallel or perpendicular to the easy $c-$axis) from the measured torque
$\Gamma$ is analytical and unambiguous. The applied magnetic field creates a
torque, which at equilibrium, is compensated by a mechanical drawback one. In
a longitudinal field the system of non-linear equations linearizes. To find a
hysteresis loop we first perform a two-parameter linear fit ($u$ and $v$) of
the saturation region with $B_0/\Gamma=u+vB_0$ and then we calculate the
normalized magnetic moment $m=M_\parallel/M_S=u/(B_0/\Gamma-vB_0)$. In the
transverse case, $M_\perp$ is, to a good enough approximation, given by
$M_\perp=\Gamma/B_0$.

\begin{figure}
\includegraphics[width=8.1cm]{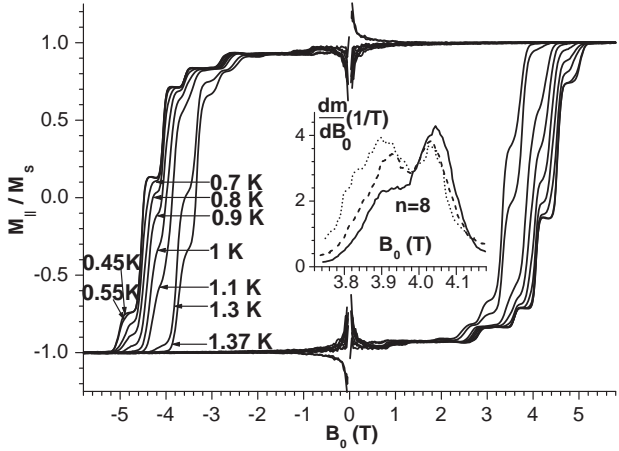}
\caption{Hysteresis loops of Mn$_{12}$-ac, derived from torque measurements
with an applied field along the $c-$axis. Below $\approx0.7$~K, the curves are
temperature-independent, indicating quantum tunneling without thermal
activation. Transitions widths result from the distribution of dipolar fields.
Inset: $dm/dB_0$ where $m=M_\parallel/M_s$  for the resonance $n=8$ at three
temperatures: 0.90 K (dots), 0.95 K (dashed) and 1.00 K (continuous).   The
resonance is split in two: tunneling from the ground-state $m=-10$ and  from
the first excited state $m=9$. Note the change of the height of each peak,
when varying the temperature.  Similar behavior was found for all the other
resonances.} \label{fig1}
\end{figure}

\begin{figure}
\includegraphics[width=8.1cm]{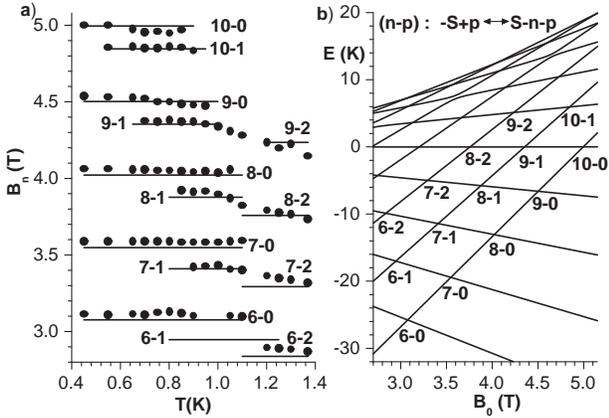}
\caption{(a): Measured positions of peak maxima \emph{vs.} temperature.
Horizontal lines indicate the calculated crossing fields. (b): Energy levels
spectrum calculated by exact diagonalization of the Hamiltonian  $H =
-DS_{z}^2 -BS_{z}^4-C(S_{+}^4 + S_{-}^4) - g\mu_BS_zH_z$ for a spin $S=10$ and
the parameters  $D=0.56$~K, $B=1.1$~mK, $C=3\times10^{-2}$~mK obtained in
EPR. The term $B$ shifts the resonances belonging to same $n$
and different $m$ to lower fields when $m$ decreases.}
\label{fig2}
\end{figure}

Fig.~\ref{fig1} shows several staircase-like hysteresis loops in a
longitudinal field, below 1.3~K. The angle between the applied field and the
easy axis, due to some inherent  misalignment in the experimental set-up, was
$\lesssim2.5^{\circ}$. The corresponding  transverse field component, smaller
than 0.2~T, is almost not affecting the transitions with $2m-n$ a multiple of
4 ($m$ is the $S_z$ eigenvalue), because of the leading $C$ term in the
transverse anisotropy (see the caption of Fig.~\ref{fig2}b) or with $n\le6$.
The observed $B_n$ are independent of temperature (Fig.~\ref{fig2}), an effect
which was also observed in~\cite{Peren} but below $\approx0.7$~K only. This
suggests that tunneling takes place from the metastable state $m=-10$ to final
states $m=10-n$ ($n=6$ to 10). As shown in Fig.~\ref{fig1} inset for $n=8$,
the $B_n$ are determined accurately enough to see that each resonance is
constituted of several temperature-independent peaks (resonance groups). Their
widths $\approx0.15$~T are close to the one observed at high temperature and
come from the distribution of dipolar fields~\cite{Luc,jmmm200}. A comparison
between the measured and calculated $B_n$ makes their interpretation obvious.
The calculated $B_n$ are given by level crossings in Fig.~\ref{fig2}b (the
Hamiltonian, given in the caption of  Fig.~\ref{fig2}b, was found by EPR
experiments~\cite{BarraPRB}). Intercepts of the $m=S=-10$ state with
$m=3$, 2, 1 and 0, give the $B_n$ indicated by horizontal lines $n$-0 in
Fig.~\ref{fig2}a. Similar intercepts of the first excited level $m=-9$ with
$m=2$, 1, 0 and -1, are also given. This shows unambiguously that tunneling
occurs, when increasing temperature, first from $m=-10$, then from the first
excited state $m=-9$, then from $m=-8$, $\dots$ Tunneling transitions  with
smaller $m$  are closer and closer and may appear to be temperature-dependent
if the field resolution is not good enough. Regarding peak intensities, the
area of deconvoluted Lorentzian peaks in each resonance group, characterized
by the same index $n$, determines the fraction of tunneling events taking
place from the ground-state (index $p=0$), from the first excited state
($p=1$), $\dots$ One can define an experimental cross-over temperature
$T_{c,n}$ for which the two peaks ($n$-0 and $n$-1) have the same area; for
example, for $n=8$ we found 0.95~K. Accounting for the proportionality of the
tunneling rate with $\Delta^2$, where $\Delta$  is the tunnel splitting
~\cite{ProkStamp}, this crossover can be written $kT_{c,n}\approx
E_{0,1,n}/2\ln(\Delta_{n-1}/\Delta_{n-0})$,  where $E_{0,1,n}$ is the
separation between the ground-state and the first excited state, near the
$n^{th}$ crossing. For the resonances with $n=8$ and $E_{0,1,8}\approx7.3$~K,
calculated by exact diagonalization, we get
$\ln(\Delta_{8-1}/\Delta_{8-0})=3.88$ corresponding to a transverse field of
$\approx0.31$~T (a misalignement of $\approx4.5^\circ$). Obviously, such a
transverse field is larger than what is expected from our experimental set-up.
Similar discrepencies are also observed with other resonances.  The important
effect of small transverse fields is not sufficient to give an overall
agreement, in particular when $2m-n$ is not a multiple of 4 and when $n\ge 6$.
Finally, early observation of a crossover temperature of $\approx 2$~K in zero
field ~\cite{BBjmmm95,LucLowT,BBjmmm98} is now confirmed by the
extrapolation of $T_{c,n}$ (Fig.~\ref{fig2}a), giving $T_{c,0}\approx1.7$~K
(the  same type of extrapolation shows that the continuous process of
ground-state to  activated tunneling up to the top of the barrier spreads
between $\approx1.7$ and $\approx2.4$ K). However, this requires a ground-state
splitting $\approx3\times10^{-10}$~K  instead of the $\approx10^{-11}$~K as
calculated in~\cite{jmmm200}. This zero-field crossover   All that suggests a
lack of knowledge of the transverse spin operator coefficients of  the
Hamiltonian used for Mn$_{12}$-ac (Fig.2b). Besides the fact  that all crystal
field parameters  of order larger than four are unknown, we must mention that
all operators with odd order are missing. The $S_4$ point symmetry of
Mn$_{12}$-ac~\cite{Lis} leads to the following Hamiltonian~\cite{Kahan}:
$H=B_{2}^{0}O_{2}^0+B_{3}^{2}O_{3}^2+
B_{4}^{0}O_{4}^{0}+B_{4}^{4}O_{4}^4+B_{5}^{2}O_{5}^2+B_{6}^{0}O_{6}^0+
B_{6}^{4}O_{6}^4+\dots$ where the $O_{l}^m$ are the Steven's equivalent
operators~\cite{Stevens}. The complete  Hamiltonian should also contain
magnetic field and the effect of Dzyaloshinsky-Moriya
terms~\cite{BBjmmm98,SeatleNotes,Dobro}. Interestingly, the new odd terms have
no diagonal contribution, which is consistent with  the good agreement we
obtain regarding the resonance positions. This also shows that diagonal terms
of   order $\ge 6$ are negligible.

\begin{figure}
\includegraphics[width=8.1cm]{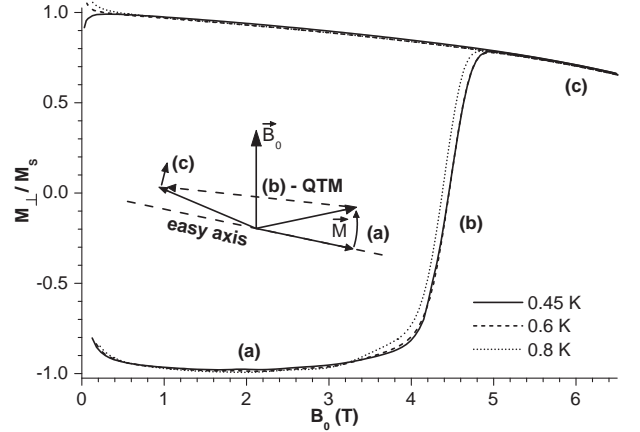}
\caption{Half hysteresis loop in transverse field at three different
temperatures. $M_\perp$  is given by the measured torque divided by field.
The crystal was first saturated in a large negative field. Inset: the geometry
of the experiment.}
\label{fig3} \end{figure}

Let us now consider the case when the field is applied perpendicularly to the
easy axis. The symmetry of the double well is preserved allowing to observe
tunneling between the ground-states $|S,+S>$ and $|S,-S>$. The sample was
first saturated along the $c-$axis of the crystal. Then, the field was set to
zero and the sample rotated to $90^{\circ}\pm1^{\circ}$. During the
experiments, the applied field $B_0$ was varied between 0 and 5~T with a
transverse component $B_T$ nearly equal to $B_0$, and a longitudinal one $B_L$
varying between 0 and $\approx-0.07$~T. When $B_L\approx-0.04$~T $\approx$ the
Lorentz internal field of Mn$_{12}$-ac, the internal field is canceled.
Nevertheless, level widths  being  $\approx0.15$~T, the spin-up and spin-down
density of states were in coincidence during the whole variation of the 
applied field. Hysteresis loops of the transverse magnetization are plotted
Fig.\ref{fig3}. They show only one jump and are nearly independent of
temperature, indicating that spin reversals take place by single coherent
ground-state tunneling events. This loop can be understood easily. In
transverse fields smaller than $\approx4$~T, tunneling rate $\Gamma_0$ between
symmetrical ground-states is too small and nothing happens. The tunneling gap
increasing almost exponentially  with $B_T$  \cite{Korenblit}, $\Gamma_0$
becomes rapidly very large, and when $B_T\approx4.5$~T the spins tunnel to the
other well during the time of the experiment; due to a small misalignment from
the ideal transverse geometry, as shown in Fig.~\ref{fig3} insert, after
tunneling the molecules are kept within the final well (the sample is again
saturated). The observation of tunneling requires  $\Gamma_0^{-1}$ to fall in
the experimental window $1-10^3$~s. Such values for the tunneling rate, which
depends  on the level splitting and the hyperfine field
fluctuations~\cite{ProkStamp}, are realistic according to NMR results on
Mn$_{12}$~\cite{Goto} and for tunnel splittings calculated between 4 and
4.5~T. 

\begin{figure}
\includegraphics[width=8.1cm]{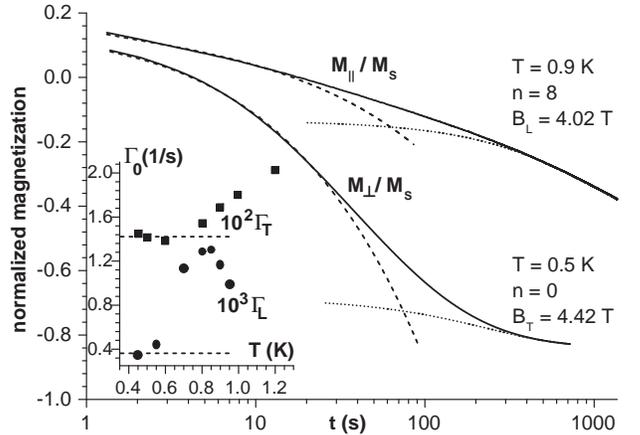}
\caption{Relaxation curves in a transverse (longitudinal) field of 4.42~T
(4.02~T) for the resonance $n=0$ at $T=0.5$~K ($n=8$ at $T=0.9$~K). In both
cases the square root regime (dashed) is observed only at short timescales,
then the relaxation becomes exponential (dotted). Note that experiments are
done far from saturation. Inset: temperature dependence of the square root
tunneling rate in transverse (squares-$\Gamma_{T}$) and longitudinal field
(circles-$\Gamma_{L}$).}
\label{fig4}
\end{figure}

Magnetic relaxation experiments give direct access to tunneling rates. An
example is given Fig.~\ref{fig4}, in longitudinal and transverse applied
fields. In both cases and below 1.2 K, the relaxation is a square root of
time at  short times: $M(0)-M(t)=M_s\sqrt{\Gamma_{sqrt} t}$. This law,
already observed in Fe$_8$~\cite{Ohm} or Mn$_{12}$-ac above 1.5
K~\cite{LucLowT}, was predicted  by Prokof'ev and Stamp~\cite{ProkStamp}. It
is characteristic of quantum tunneling in the presence of dipolar interactions
and fast hyperfine fluctuations and ends for times
$\approx\Gamma_{sqrt}^{-1}$ due to the building of multimolecule spin
correlations. At longer times more and more important correlations, should in
principle lead to non-exponential relaxation (see Fig. 1 in
\cite{ProkStamp}). However, this is not what we observe in Fig.~\ref{fig4},
where the quantum relaxation regime, succeeding the square root one at long
times, is exponential. We believe that this is because during the smooth
transition between the square root and exponential regimes, couplings to the
environment (spins, phonons$\dots$) have enough time to erase multimolecule
spin correlations, an effect which could be called ``environmental annealing".
 The crossover time $t_{c}$ (when the dashed fit intercepts the dotted fit in
Fig.~\ref{fig4}), depends on many parameters and will be analysed in further
studies. Regarding the first regime and for our present experimental
conditions  (sample in good contact with the cryostat), $\Gamma_{T,L}(T)^{-1}$
is always of the order of $t_{c}$, showing that fast recovery has the same
origin as fast spin relaxation, i.e. not too small tunnel splittings and large
hyperfine field fluctuations~\cite{ProkStamp,JLTP96}. As shown above
spin-phonon transitions also contribute to $\Gamma_{T,L}$ and thus to recovery
near $T_{c,n}$ by adding new, thermally excitated channels, with larger tunnel
splittings and faster relaxation rates.

In Fig.~\ref{fig4} inset are shown the square-root relaxation rates plotted
\emph{vs.} temperature, measured at the 8-0 resonance (Fig.~\ref{fig2}) in
longitudinal geometry (full circles-$\Gamma_{L}$) and in transverse geometry
for the 0-0 resonance (full squares-$\Gamma_{T}$). As expected from the first
part of this paper, $\Gamma_{L}$ increases above  $\approx0.6$~K, then it goes
through a maximum and then decreases. The increase of $\Gamma_{L}(T)$ is due
to the proximity of the resonance 8-1. The field separation between this
resonance at 3.90~T and the resonance 8-0 at which $\Gamma_{L}(T)$ is measured
at 4.02~T, being of the order of level broadening by dipolar interactions,
spin-phonon transitions between them are allowed. This is what we observe and
constitutes a clear example of  phonon-assisted tunneling in the quantum
regime. Regarding peak intensities,  near equilibrium the rate increases like
$\Gamma_L(T)\approx\Gamma_{L,8-0}+\Gamma_{L,8-1}\exp(-E_{0,1,8}/kT)$. As an
example, for $T=0.7$~K and $\Gamma_{L,8-0}=\Gamma_L(0.55)$, we obtain
$\ln(\Delta_{8-1}/\Delta_{8-0})\approx5.35$ too large to be explained
numerically. As we already mentioned, new terms in the spin Hamiltonian of
Mn$_{12}$ could lead to a better agreement. The decreasing part of
$\Gamma_{L}(T)$ simply comes from a decreasing of the number of states
available for phonon-assisted tunneling, due to the magnetization
decrease when the field sweeps across the 8-1 resonance. Like $\Gamma_{L}$,
the square-root relaxation rate $\Gamma_{T}$, measured in transverse geometry,
is independent of temperature at lowest temperatures, as expected for ground
state tunneling. Above $\approx0.7$~K the tunneling rate increases with
temperature. However this  is a very smooth effect which cannot be due to
activated tunneling on levels $m=9$ and $m=8$.  It is rather due to
tunneling assisted by spin-phonon transitions  between $m\approx\pm10$
states, split by dipolar fields (and integrated over the spin-up  and
spin-down distributions). This assumption is supported by the fact that the
corresponding energy scale of $\approx0.5$~K for increasing $\Gamma_T$ is
close to the dipolar energy ($g\mu_BSB_{dipolar}\approx0.5$~K). This is
another example of phonon-assisted tunneling in the quantum regime.

In conclusion, we have shown that in Mn$_{12}$-ac with a longitudinal field
($n\ne0$), the transition between coherent tunneling on the ground-state and on
upper states (with thermal activation) takes place, level after level, when
they come into coincidence in an applied sweeping field. This is an obvious
consequence of the $4^{th}$ order anisotropy term $BS_{z}^4$, which prevents
the spin-up and spin-down level schemes coinciding. In any case, the
experiments show a continuous transition, even in a constant field, contrary to
Ref.~\cite{Garanin}. More interestingly, phonon-assisted tunneling is
evidenced between $m=-10$ and $m=10$ splitted by dipolar fieds in the
transverse case or between $m=-10$ and $m=1$ in the longitudinal one. 
``Annealing" of multimolecule spin correlations in the quantum regime result
from spin relaxation itself and from spin-phonon transitions near the
crossover to excited states. Such an exponential phonon-mediated relaxation
cross-over was predicted for a single-molecule relaxation~\cite{JLTP96}, but
no theoretical work exists which takes into account dipolar interactions. It
would be of great interest to see a theoretical treatement of the long time
quantum relaxation in the presence of phonons and dipolar interactions.
Finally we suggest not to forget that spin operators with odd powers are
allowed by the $S_4$ symmetry of the Mn$_{12}$-ac molecule.

We thank I.Tupitsyn and S. Miyashita, S. Maegawa and T. Goto for discussions.

\end{document}